# Prediction of superconducting properties of materials based on machine learning models

**Jie Hu[1], Yongquan Jiang[*2], Yang Yan[2], Houchen Zuo[1]**

(1 State Key Laboratory of Traction Dynamics, Southwest Jiaotong University, Chengdu 610031, Sichuan Province, China.

2 School of Computer and Artificial Intelligence, Southwest Jiaotong University Chengdu, Sichuan Province 611756, China)

**Abstract**: The application of superconducting materials is becoming more and more widespread. Traditionally, the discovery of new superconducting materials relies on the experience of experts and a large number of "trial and error" experiments, which not only increases the cost of experiments but also prolongs the period of discovering new superconducting materials. In recent years, machine learning has been increasingly applied to materials science. Based on this, this manuscript proposes the use of XGBoost model to identify superconductors; the first application of deep forest model to predict the critical temperature of superconductors; the first application of deep forest to predict the band gap of materials; and application of a new sub-network model to predict the Fermi energy level of materials. Compared with our known similar literature, all the above algorithms reach state-of-the-art. Finally, this manuscript uses the above models to search the COD public dataset and identify 50 candidate superconducting materials with possible critical temperature greater than 90 K.

## 1 Introduction

Machine learning is increasingly used in computing material properties due to its absolute advantage in big data(Raccuglia et al ,2016; Avery et al ,2019) . Machine learning not only allows researchers to obtain the properties of compounds in a relatively short time, but also saves experimental costs. After the discovery of superconducting materials, scientists have continued to research superconductors because of their promising applications in power transmission, high currents, electronics, and antimagnetism, and have successively discovered some superconductors with high critical temperatures. For example, the critical temperatures of sulfur cyanide and lanthanide have been found to exceed 200 K under pressure conditions exceeding 150 GPa (A et al ,2015; M et al ,2019); at normal temperature and pressure, scientists have found copper to have a higher critical temperature(Bednorz et al ,1986). However, all of the above methods of discovering superconducting materials require the "experience" of experts and are obtained through extensive experiments. There is a systematic underestimation of the band gap, Fermi energy level, and other properties of materials in standard DFT calculations(Chan et al ,2015). Therefore, the use of new big data methods for predicting the properties of materials holds promise for solving this problem.

The main contributions of this research work are as follows.

- proposed to use XGBoost model to identify superconducting materials, and improved the identification accuracy.
- A sub-network neural network model is proposed for predicting the Fermi energy level of materials, and the prediction accuracy is improved.
- proposed the use of deep forest model to predict the critical temperature of superconducting materials, material band gap, and improve the prediction accuracy.

The rest of the manuscript structure is organized as follows: Sect. 2 represents the short literature survey, Sect. 3 represents the proposed subnetwork neural network model, and Sect. 4 illustrates the experimental results and analysis. Finally, the research work is concluded in Sect. 5.

# 2 Related work

The application of machine learning in materials science is gradually increasing due to the great success of machine learning in image processing(Zhang et al ,2019), natural language processing, etc. In superconducting materials, (Hutcheon et al ,2020) applied a machine learning approach to identify superconducting hydrides by identifying and screening stable candidates through structure search, and then performed electron-phonon calculations to obtain the critical temperatures of the candidates, predicting critical temperatures up to 115 K for RbH at 50 GPa and 90 K for CsH at 100 GPa The Gaussian process regression (GPR) model was developed by (Zhang et al ,2021). to reveal the relationship between the process parameters and the superconducting transition temperature of BiPbSrCaCuOF superconductors, which has a high stability and accuracy and has a promising use; (Owolabi et al ,2016) developed a computational intelligence model with lattice parameters as descriptors by support vector regression (SVR) (CIM) for 31 different YBCO superconductors to estimate Tc, and the estimated superconducting transition temperatures are in good agreement with the experimental values with high accuracy. The developed CIM can estimate the temperature of any processed YBCO superconductor quickly and accurately without any complex equipment; (Alizade et al ,2018). used the Debye temperature and critical transition temperature as descriptors to build a machine learning model to predict the electron-phonon coupling constants of 28 elements by cross-validation technique with an accuracy of 0.88 ; (Lee et al ,2021). used supervised learning algorithms and convolutional neural networks to successfully identify the presence of the Majorana zero mode, a hallmark of topological superconductivity, and demonstrated that the neural network model worked best; (Roter et al ,2020). used chemical descriptors of superconductors as input to a machine learning model and obtained a 0.93 coefficient of determination, and also predicted several other promising new superconductors based on this model; (Zhang et al ,2021). developed a Gaussian process regression model to predict the critical temperature of doped iron-based superconductors based on structural and topological parameters (including lattice constants, volume and bonding parameters topological index H31), which is stable and accurate and helps to estimate the critical temperature quickly. (Konno et al ,2021). proposed a method called "periodic table reading", which represents the periodic table in a deep learning way, learning to read the periodic table and learning elemental laws to discover new superconductors beyond the training data, and the model obtained the $R^2$value of the predicted material Tc as 0.92 in the superconductor database. (Stanev et al ,2018). modeled the critical temperature (Tc) of more than 12,000 known superconductors with the SuperCon database, first classifying the materials into two categories based on their Tc values above and below 10 K, and training a classification model to predict this label, which used coarse-grained features based on chemical composition only, which predicted up to 92% accuracy ,and then regression models were built to predict Tc values for Cu-based, Fe-based and low Tc compounds, respectively, these models also demonstrated good performance with the learned predictors that can provide potential insights into the mechanisms behind superconductivity in different material families by combining classification and regression models into a single integrated pipeline and used to search the entire inorganic crystal structure database (ICSD) for potential novel superconductors, resulting in the identification of 30 non-Cu-based and non-Fe oxides as candidate materials. (Zeng et al ,2019). developed an atomic table convolutional neural network (ATCNN) that requires only elemental composition to learn relevant features from its own construction, and the accuracy of the model exceeds the results of standard DFT calculations. Using the trained model, more than 20 potential compounds with high critical temperatures were screened from the existing database. (Jha et al ,2018). proposed a deep learning model, Elemnet, which automatically captures the physical and chemical interactions and similarities between different elements, thus enabling the prediction of material properties with higher accuracy and speed. Bandgap is a very important property of materials, and in predicting material bandgap, (Bart et al ,2019). trained a machine learning model using a set of datasets consisting of crystal structure and DFT bandgap from the Organic Materials Database (OMDB) with a model MAE as low as 0.388 eV. (Yang et al ,2019). trained and tested the machine learning model by using the bandgap and bandshift datasets of GaN type nitrides accurately calculated by combining the screening mixture function of HSE and DFT-PBE, the results show that support vector regression with radial kernel performs best in predicting band gap and band offset with a prediction MAE of 0.298eV and 0.183eV, respectively. (Ya et al ,2018). developed a machine learning model based on material composition that can accurately predict the band gap of inorganic solids. Experimental results show that the band gap predicted by machine learning is significantly closer to the experimentally reported values than the band gap calculated by DFT (PBE). The Fermi energy level energies of materials are critical for the design of conducting materials, heterogeneous structures and devices. Traditional methods for calculating Fermi energy levels include density flooding theory and DFT methods, which are not only computationally intensive but also costly, so machine learning models can be used to predict the Fermi energy levels of materials. (Benyamin et al ,2022). used machine learning methods to demonstrate that the main determinant of Fermi energy levels is the ionic charge, independent of its own structure.

In this manuscript, we construct different machine learning models to predict whether the material is superconducting or not, superconducting critical temperature, band gap, and Fermi energy level.

# 3 Data and proposed methodology

## 3.1 Data sources

In this manuscript, the critical temperature data of superconducting materials are obtained from the SuperCon public dataset, containing 21263 records(Benyamin et al ,2011). In this manuscript, we extracted 9399 high-energy stable insulators with DFT band gaps greater than 0.1 eV from the Materials Project database as a non-superconductor dataset. We also extracted 3896 Fermi energy level (Ef) data from the Open Quantum Database (OQMD) (Saal et al ,2013). , as well as 5886 compounds with forbidden band width (Eg) from previous literature (Zhuo et al ,2018).

The extracted data on the critical temperature of superconducting materials are preprocessed. Materials with the same composition usually have different Tc values due to the different experimental conditions under. For example, $H_2S$, which is not a superconductor under ambient conditions, has a particularly high Tc under high pressure(Saulius et al ,2012).There are two different Tc values for $H_2S$ in the database, 185 K and 60 K, respectively. To avoid confusion, for such compounds with multiple Tc values, if the maximum value exceeds twice the minimum value, the data is deleted. Otherwise, the average value is taken as the Tc value of the material and the duplicates are deleted. Unproven superconducting materials, such as $HWO_3$ , are deleted as well. Data with elemental coefficients greater than 50 in the chemical formula, such as $Hg_{1234}$ $O_{10+z}$ and data with uncertain oxygen content Determined data such as $Yb_{16}BaCuO_{27}$ are excluded. In the dataset, compounds Hg, $MgB_2$ , FeSe and $YBa_2Cu_3O_7$ are typical representatives of elemental superconductors, conventional BCS superconductors, iron-based superconductors and copper-based superconductors, respectively. They are used to test the generalization ability of the ATCNN model. Before segmenting the training and test sets, these compounds, including Hg, $Mg_xB_y$, $Fe_xSe_y$ and $YBa_2Cu_xO_y$ (where x and y denote the content of the corresponding elements) are removed from the cleaned-up dataset. The collated dataset has a total of 13598 superconductors. The data of non-superconductor materials as well as forbidden band widths and Fermi energy levels are not further processed because they do not contain duplicate data and were measured in the same environment.

### 3.2 Machine learning models

For the problem of identifying superconducting materials, this manuscript uses the XGBoost classification model; for the problem of predicting the critical temperature and material band gap of superconducting materials, this manuscript uses the deep forest model, which has better performance than other decision tree integrated learning models, as well as fewer hyperparameters and less tuning parameters and higher training efficiency.

For the Fermi energy level of the material, the neural network structure named sub-network is shown in Figure 3. In this manuscript, there are a total of six attributes, namely, atomic frequency, electronegativity, electron affinity, first ionization energy, electron affinity energy, and number of outermost electrons, all of which are commonly and easily obtained features. Each feature is extracted with one network as shown in the left part of Figure 1. After repeated experiments, we finally determined the number of layers of each network structure and the number of neurons. There are 6 fully-connected layers and the number of neurons in each layer is 60, 100, 200, 100, 50, 16, respectively, which is the blue part in Figure 1. The fully-connected layer is followed by the Dropout layer in orange. Dropout can then dormant some neurons, reduce the number of model parameters, thus speed up the model training and also improve the generalization performance of the model. After the Dropout layer, the batch normalization layer is added, which is the yellow part in Figure 1. The main role of this layer is to ensure that the distribution of the data in each layer of the network does not change significantly with the change of parameters, thus avoiding the phenomenon of internal covariate shift and ensuring a more stable training process. The last fully-connected layer uses the Linear activation function as the mapping function, which is the green part in Figure 1. The output of the six networks are connected to a new network. The network mainly consists of fully-connected layer, Dropout layer and batch normalization layer, 4 cells in total. After repeated experiments, the number of neurons in each cell is finally determined to be 80, 64, 32, and 8 respectively, and the activation function of each layer is the Tanh function. The final mapping function uses Linear to map the matrix to a specific value, thus achieving the regression of the whole sub-network.

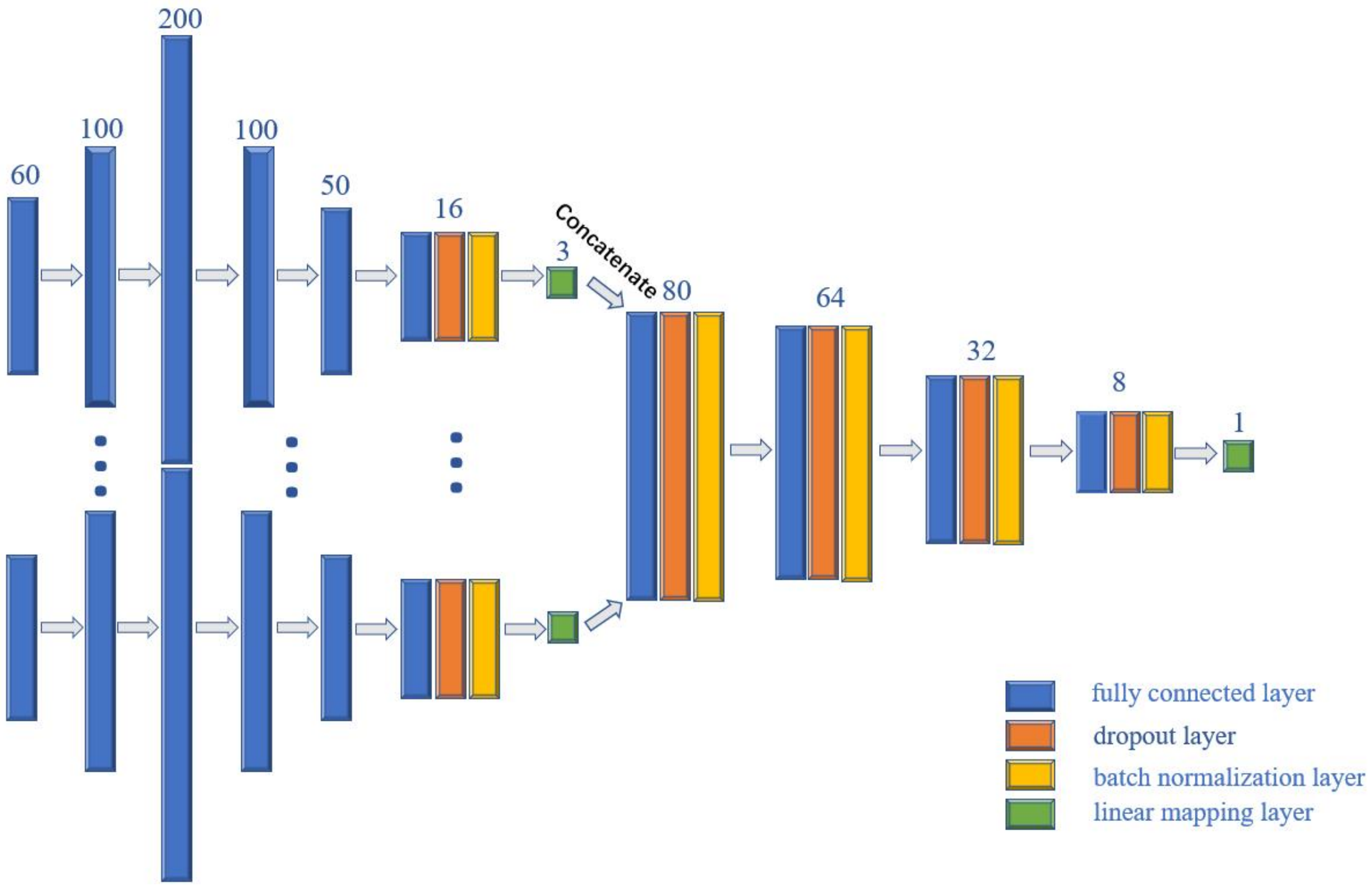

Figure 1 Schematic diagram of the sub-network

To avoid the occurrence of overfitting, Dropout is used with the dropout rate set to 0.1. The training process is terminated after 400 epoches, because the loss hardly decreases after that. The loss graph is shown in Figure 2(c).

# 4 Experiments and results

Each dataset is randomly divided into a training set (80%) and a test set (20%), and the experimental results of all models below are obtained using the same dataset. For the dataset identifying superconductors and non-superconductors, the output label of superconductors is set to 1 and the output label of 9399 insulators is set to 0. After training with the XGBoost classification model, the results are shown in Figure 1, and the AUC and TPR in the test set are 0.98 and 98.48% respectively. The results compared with other models are shown in Table 1, which shows that the AUC and TPR of the XGBoost model are better than those of the ATCNN model (Zeng et al ,2019). , the only similar literature to our knowledge.

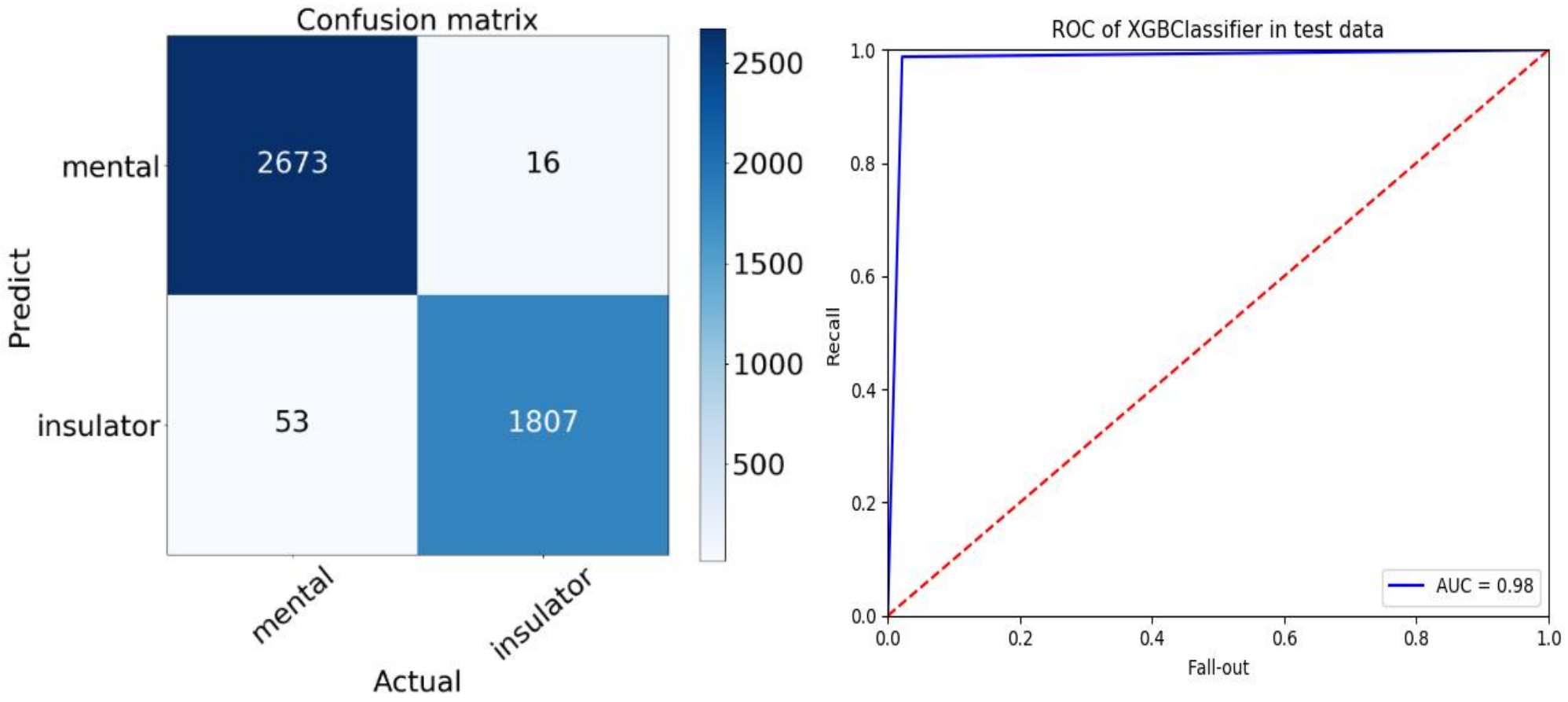

(a) (b)

Fig1 XGBoost Classifier

Table 1 Experimental results of different classification models

| Models | AUC | TPR (%) |
|---|---|---|
| GBDT | 0.96 | 96.47 |
| ATCNN(Zeng et al ,2019). | 0.97 | 93.78 |
| SVR | 0.97 | 91.9 |
| DecisionTree | **0.98** | 97.84 |
| RandomForest(Saal et al ,2013). | **0.98** | 98.23 |
| XGBoost | **0.98** | **98.48** |

For the collated superconducting critical temperature dataset, this manuscript proposes to use the atomic frequency as the input of the deep learning model and the deep learning model we chose is deep forest. The experimental results are shown in Fig. 2(a), and in the test set, the value of R2, MAE and RMSE are 0.944, 4.04, and 7.51 respectively. Using the same training set and test set, the comparison results with similar literature are shown in Table 2 with the bolded indicating the best result in that column. In this manuscript, all algorithms in Table 2 are implemented and tested in the same hardware and software environment.

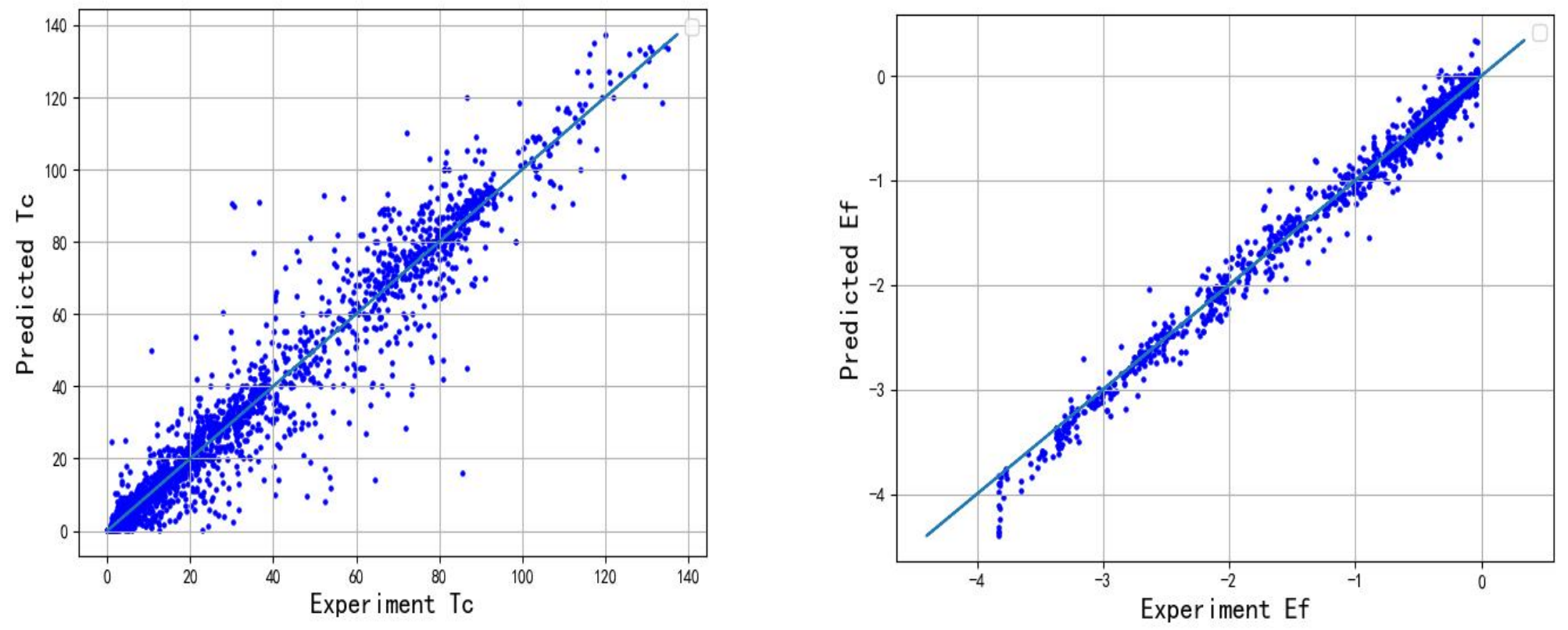

(a) (a) Scatter diagram of critical temperature (b) Scatter diagram of Fermi energy level

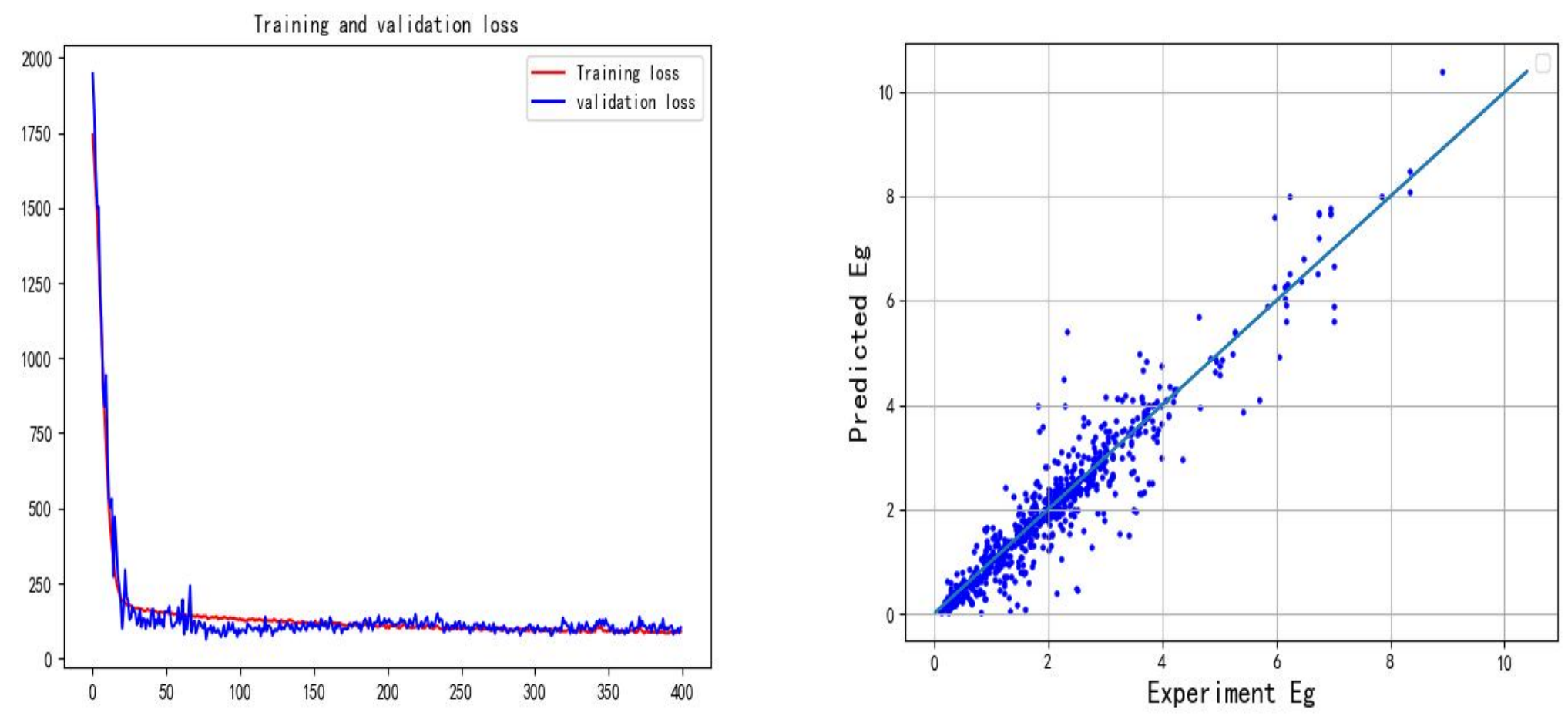

(c) Sub-network loss plot (d) Band gap scatter plot

Figure 2 Graph of prediction results

Table 2 Summary results of the critical temperature prediction experiments

| Network Model | $R^2$ (%) | MAE | RMSE |
|---|---|---|---|
| SVM | 71.4% | 11.36 | 17.06 |
| K Nearby | 89.1% | 5.88 | 10.40 |
| DecisionTree | 88.8% | 5.40 | 10.50 |
| GBDT | 85.2% | 8.33 | 12.16 |
| ExtraTree | 89.2% | 5.32 | 10.40 |
| 1DCNN | 89.1% | 6.25 | 10.54 |
| Artificial Neural Networks | 89.6% | 6.15 | 10.64 |
| XGBoost | 91.6% | 5.71 | 9.34 |
| Bagging | 91.8% | 4.80 | 9.07 |
| ATCNN(Zeng et al ,2019). | 89.1% | 5.88 | 10.40 |
| Subnetwork | 92.0% | 5.41 | 8.78 |
| Random Forest(Saal et al ,2013). | 92.8% | 4.58 | 8.53 |
| Deep Forest | **94.5%** | **4.04** | **7.51** |

Table 2 shows that the value of R2, MAE, and RMSE of the test set of the deep forest model are 94.5%, 4.04, and 7.51 respectively, which are optimal in the published literature.

The scatter plot of the material Fermi energy level is shown in Fig. 2(b) and the final results obtained using different models are shown in Table 3 with the bolded indicating the best value in that column. From Table 3, it can be seen that the sub-network model can achieve more accurate predictions of the Fermi energy levels. For the Fermi energy level dataset, the MAE of ExtraTree and the subnetwork model are both 0.1, which is lower than the MAE of DFT in literature (Kirklin et al ,2015).and the 0.15 of ElemNet model in literature (Dipendra et al ,2018). This also means that the accuracy of the model used in this manuscript exceeds the DFT calculation. The $R^2$, MAE, and RMSE value of the sub-network model can also be seen in Table 3 at 98.4%, 0.10, and 0.14 respectively.

Table 3 Summary results of Fermi energy level prediction experiments

| Network Model | $R^2$ (%) | MAE | RMSE |
|---|---|---|---|
| 1DCNN | 95.2% | 0.17 | 0.25 |
| K Nearby | 86.7% | 0.26 | 0.40 |
| SVM(Owolabi et al ,2016). | 88.7% | 0.24 | 0.38 |
| ExtraTree | 96.6% | **0.10** | 0.20 |
| XGBoost | 93.8% | 0.20 | 0.27 |
| ElemNet(Dipendra et al ,2018). | - | 0.15 | - |
| ATCNN(Zeng et al ,2019). | 96.5% | 0.13 | 0.21 |
| Random Forest(Stanev et al ,2018). | 95.5% | 0.13 | 0.23 |
| DFT(Kirklin et al ,2015). | - | 0.136~0.81 | - |
| Deep Forest | 97.7% | 0.11 | 0.17 |
| Subnetwork | **98.4%** | **0.10** | **0.14** |

In this manuscript, the atomic frequency number is used as the input for the deep learning model. The deep forest model is chosen to predict the forbidden band width of the material. In the test set, the MAE, RMSE, and R2 of the deep forest model are 0.27, 0.44, and 0.917 respectively. The corresponding scatter plot is shown in Fig. 2(d). The results are shown in Table 4 with the bolded indicating the best value of the column.

Table 4 Summary results of bandgap prediction experiments

| Network Model | $R^2$ | MAE | RMSE |
|---|---|---|---|
| SVM(Owolabi et al ,2016). | 55.5% | 0.59 | 0.56 |
| ExtraTree | 61.7% | 0.44 | 0.83 |
| 1DCNN | 64.9% | 0.52 | 0.82 |
| K Nearby | 69.7% | 0.45 | 0.79 |
| ATCNN(Zeng et al ,2019). | 81.4% | 0.35 | 0.63 |
| Random Forest(Stanev et al ,2018). | 81.1% | 0.34 | 0.64 |
| XGBoost | 81.8% | 0.39 | 0.64 |
| Bagging | 82.5% | 0.34 | 0.64 |
| Subnetwork | 86.6% | 0.34 | 0.55 |
| CGCNN(Xie et al ,2018). | - | 0.388 | - |
| DFT(Kirklin et al ,2015). | - | 0.6 | - |
| Deep Forest | **91.7%** | **0.27** | **0.44** |

As shown in Table 4, the MAE, RMSE, and R2 of the deep forest model all outperform other machine learning models such as ATCNN, CGCNN, and SVM. It also outperforms the DFT calculation method, which shows that the accuracy of the deep forest model has surpassed the accuracy of the DFT calculation method.

In order to verify the effectiveness of the depth forest model in predicting the critical temperature of superconductivity, four materials were extracted from (Zhuo et al ,2018)., all of which have measured critical temperature values. The values predicted using the depth forest model in this manuscript were compared with literature (Zhuo et al ,2018). and the results are shown in Table 5, with the bolded indicating the best value in the row. Table 5 shows that the prediction error of the deep forest model for $CaBi_2$ is much lower than that of literature (Zhuo et al ,2018). and the prediction errors of the remaining materials are within 3%.

Table 5 Critical temperature verification results

| Materials | Measured Tc value (K) | The literature predicts that Tc values (K) (Konno et al ,2021). | Deep forest model predicts Tc values (K) |
|---|---|---|---|
| $CaBi_2$ | 2 | 14.85 (642%) | **5.91** (195%) |
| $HfV_4$ Zr | 10 | 10.17 (1.7%) | **9.92** (0.8%) |
| $Au_{0.5}$ $Nb_3$ $Pt_{0.5}$ | 10 | **10.13** (1.3%) | 10.30 (3%) |
| $Hf_{0.5}$ $Nb_{0.2}$ $V_2$ $Zr_{0.3}$ | 10 | **10.11** (1.1%) | 9.83 (1.7%) |

To further verify the practicality of the model, in this manuscript, for the 100,000 materials in the COD database(Saulius et al ,2012)., each material is first identified using the XGBoost model. If it is a superconducting material, the critical temperature value is then predicted using the deep forest model. Finally the candidate superconducting materials with a critical temperature greater than 90 K are screened with a total of 50 materials satisfying the conditions collected. The detailed results are shown in the Appendix.

# 5 Conclusion

(1) Random Forest, XGBoost and Decision Tree models can all be used as models for identifying superconducting materials, among which XGBoost has the highest accuracy. This model outperforms the classification models in ATCNN (Zeng et al ,2019).

(2) For the prediction of superconducting critical temperature and forbidden band width of the material, the deep forest model has the best determination coefficient $R^2$, mean absolute error and root mean square error compared with similar literature.

(3) For the Fermi energy level of the material, the determination coefficient $R^2$, the mean absolute error, and the root mean square error of the subnetwork model we proposed are optimal compared with similar literature.

**Funding**

The research work was supported by the National Natural Science Foundation of China (No.61976247) and the Fundamental Research Funds for the Central Universities Universities (No.2682021ZTPY110).

# Appendix

| Material Name | Predicted Critical temperature |
|---|---|
| Ba1.9Ca1.9Cu3O9Tl1.1 | 106.39075 |
| Bi2.1Ca2Cu3O10Sr1.9 | 102.2544 |
| Ba2Ca0.9Cu2.14O8Tl1.96 | 98.941975 |
| Ba2Cu3ErO6.98 | 90.841325 |
| Ba2Ca2Cu3.134O9.808Tl1.778 | 114.857725 |
| Ba2Ca1.86Cu3O9Tl1.12 | 108.329775 |
| Ba8Ca8Cu12O39Tl7 | 99.241375 |
| Ba2Ca1.95Cu3.25O9.952Tl1.66Y0.05 | 107.52595 |
| Ba4Ca4Cu6O19Tl3 | 106.2200375 |
| Ba6Ca6Cu9O29Tl5 | 101.8035625 |
| Ba2Ca1.07Cu2O8Tl1.93 | 101.905225 |
| Ba2Cu3ErO6.99 | 90.801625 |
| Ba1.95CaCu2Hg0.69O7Sr0.05V0.31 | 101.5832917 |
| Ba1.59Ca0.9Cu2Hg0.69O7Sr0.51V0.31 | 93.60211667 |
| Ba1.7CaCu2Hg0.67O7Sr0.3V0.33 | 98.56374167 |
| Ba1.33Ca0.87Cu2Hg0.69O7Sr0.8V0.31 | 91.120875 |
| Ba2Ca1.856Cu3.276O10Tl1.864 | 113.42405 |
| Au0.17Ba2Ca2Cu3Hg0.69O8.3 | 131.5497583 |
| Ba2Ca2Cu3O8.84Tl0.93 | 112.3026875 |
| Ba2Ca1.9Cu3O10.94Tl1.82 | 115.1047667 |
| Ba2Ca2Cu3Hg0.692O8.6 | 131.5497583 |
| Ba2CaCu2O8Tl1.81 | 101.761875 |
| Ba0.72Ca1.84Cu3O9Sr1.28Tl1.16 | 100.51835 |
| Ba2Ca2Cu3O8.5Tl | 113.4216125 |
| Ba2Cu3O6.958Y | 90.2204 |
| Bi2CaCu2O9.07Sr2 | 94.4891125 |
| Ba1.92Ca1.9Cu2.91O10Tl2.27 | 107.3811 |
| Ba2Ca0.93Cu2O7.86Tl1.81 | 100.6902625 |
| Ba2Ca0.93Cu2K1.12O8.46Tl0.88 | 98.5114375 |
| Ba1.9Ca0.94Cu1.9O8Tl2.26 | 93.95225 |
| Ba1.96Ca0.93Cu1.96O8.17Tl2 | 101.1992875 |
| Cr2CuO4 | 99.03175 |
| Ba2Ca0.72Cu2O8Tl2.16 | 91.5507 |
| Ba2Ca1.93Cu2.862O9Tl1.07 | 113.4243125 |
| Ba2Ca0.84Cu2O8Tl1.94 | 104.2751375 |
| Ba1.93Cu3K0.07O7Y | 90.040525 |
| Ba2CuHgO4.172 | 90.98420893 |
| Ba2CuHgO4.069 | 90.98420893 |
| Bi2Ca1.7Cu3F4O8Pb0.3Sr2 | 97.19185 |
| Bi2Ca1.7Cu3O10Pb0.3Sr2 | 94.6921 |
| Ba1.6Ca2Cu3O9.752Sr0.4Tl1.772 | 105.798625 |
| Ba1.7Ca2Cu3O9.728Sr0.3Tl1.778 | 104.9419375 |
| Bi0.3Ca3Cu4Hg0.7O10.74Sr2 | 103.4254833 |
| Bi0.3Ca1.74Cu3Hg0.7O8.91Sr2.26 | 104.1335 |
| Ba2Cu3NDyO6.9 | 90.341175 |
| Ba2Co0.03Cu2.97O6.94Y | 91.344125 |
| Ba4Ca5Cu7Hg1.44O20Re0.5 | 105.889575 |
| Ba2CaCu2Hg0.7O7.8Tl1.3 | 93.18645 |
| Ba1.66Ca0.9Cu2Hg0.7O7.4Sr0.44Tl1.3 | 92.3383 |
| Ba1.5Ca2Cu3O9.784Sr0.5Tl1.81 | 106.0555125 |